\documentclass[aps,pre,twocolumn,showpacs,superscriptaddress]{revtex4}
\usepackage{graphicx}
\usepackage{dcolumn}
\usepackage{bm}
\usepackage{amssymb}
\begin{document}
\preprint{APS}
\title{Multi-particle collision simulations of 2D one-component plasmas:\\
anomalous transport and dimensional crossovers
}
\author{Pierfrancesco Di Cintio}
\email{p.dicintio@ifac.cnr.it}
\affiliation{Consiglio Nazionale delle Ricerche, Istituto di Fisica Applicata ``Nello Carrara" via Madonna del piano 10, I-50019 Sesto Fiorentino, Italy}
\affiliation{Dipartimento di Fisica e Astronomia and CSDC, Universit\'a di Firenze, 
via G. Sansone 1, I-50019 Sesto Fiorentino, Italy}
\affiliation{Istituto Nazionale di Fisica Nucleare, Sezione di Firenze, via G. Sansone 1, I-50019 Sesto Fiorentino, Italy}
\author{Roberto Livi}
\affiliation{Dipartimento di Fisica e Astronomia and CSDC, Universit\'a di Firenze, 
via G. Sansone 1, I-50019 Sesto Fiorentino, Italy}
\affiliation{Istituto Nazionale di Fisica Nucleare, Sezione di Firenze, via G. Sansone 1, I-50019 Sesto Fiorentino, Italy}
\affiliation{Consiglio Nazionale delle Ricerche, Istituto dei Sistemi Complessi via Madonna del piano 10, I-50019 Sesto Fiorentino, Italy}
\author{Stefano Lepri}
\affiliation{Consiglio Nazionale delle Ricerche, Istituto dei Sistemi Complessi via Madonna del piano 10, I-50019 Sesto Fiorentino, Italy}
\affiliation{Istituto Nazionale di Fisica Nucleare, Sezione di Firenze, via G. Sansone 1, I-50019 Sesto Fiorentino, Italy} 
\author{Guido Ciraolo}
\affiliation{CEA, IRFM, F-13108 Saint-Paul-lez-Durance, France}
 \date{\today}
\begin{abstract}
By means of hybrid multi-particle collsion--particle-in-cell (MPC-PIC) simulations we study the dynamical scaling of energy and density correlations at equilibrium in moderately coupled 2D and quasi 1D plasmas. We find that the predictions of Nonlinear Fluctuating Hydrodynamics for the structure factors of density and energy fluctuations in 1D systems with three global conservation laws hold true also for two dimensional systems that are more extended along one of the two spatial dimensions. Moreover, from the analysis of the equilibrium energy correlators and density structure factors of both 1D and 2D neutral plasmas, we find that neglecting the contribution of the fluctuations of the vanishing self-consistent electrostatic fields overestimates the interval of frequencies over which the anomalous transport is observed. Such violations of the expected scaling in the currents
correlation are found in different regimes, hindering the observation of the asymptotic
scaling predicted by the theory.
\end{abstract}
\pacs{34.10.+x, 52.20.Hv,  52.65.-y}
\maketitle
\section{Introduction}
Many-particle systems with one or two spatial degrees of freedom $d$ often show anomalous transport properties \cite{2003PhR...377....1L,DHARREV,lepri2016thermal}. For nonlinear
lattice models, the heat conductivity coefficient $\kappa$ is found to diverge with the system size $N$ as a power-law for $d=1$ \cite{1997PhRvL..78.1896L,wang2016simulation}, and logarithmically for $d=2$ \cite{Lippi00,2012PhRvE..86d0101W}, thus leading to the breakdown of the classical Fourier law. Qualitatively, the anomalous behavior of $\kappa$ and other transport coefficients can be traced back to the constraints on the dynamics of fluctuations and collective excitations in low dimensionality, as well as to the longer {\it relaxation times} of the latter. Analytical studies based on non-linear fluctuating hydrodynamics theory (hereafter NFH) \cite{2012PhRvL.108r0601V,2014JSP...154.1191S,spohn2016fluctuating}, 
unveiled the relation between anomalous transport in anharmonic chains and the 
fluctuating Burgers/Kardar-Parisi-Zhang (hereafter KPZ) equations for the interface growth \cite{1986PhRvL..56..889K}.
\indent It is nowadays well established on theoretical and numerical grounds, that one-dimensional nonlinear systems with three conservation laws (e.g. mass, total energy and momentum) generically fall in the same KPZ universality class where $\kappa\propto N^{1/3}$ \cite{spohn2016fluctuating,2015JSP...tmp...48S}. 
This is somehow intermediate between diffusive ($\kappa\propto N^{0}$)
and ballistic (i.e., $\kappa \propto N$) transport. The latter occurs in
integrable models, e.g.  the chain of harmonic oscillators \cite{RLL67} 
and the Toda lattice  \cite{Zotos02} due to the fact that energy is transmitted through undamped propagation of eigenmodes (respectively phonons and solitons). 
More recently, it has been argued that two main
nonequilibrium universality classes, the diffusive and KPZ, are only two cases of an infinite discrete family \cite{2015PNAS..11212645P}. The members of this family can be identified by their dynamical exponent that depends on both the number of conserved quantities and on the coupling among their hydrodynamic modes.\\ 
\indent If the picture for one-dimensional systems is well developed, 
much less is known for two-dimensional systems \cite{2002cond.mat..4247G,2016EL....11324003S}. 
Here a complete NFH theory has not yet been developed and also numerical studies 
are relatively scarce. For instance, the paradigmatic 2D Ising model shows normal conduction independently on its temperature $T$ \cite{1999PhRvE..59.2783S}. 
Some numerical studies on 2D square oscillator lattices confirmed the expected logarithmic divergence of heat conductivity  \cite{Lippi00,2012PhRvE..86d0101W,Barik2007,wang2016simulation}.
Evidences of dimensional crossovers from quasi 1D to 2D scaling has been also
reported \cite{2002cond.mat..4247G,wang2016simulation}.
Another remarkable case is the Hamiltonian $XY$-model that displays a transition between logarithmically divergent and normal conductivity when increasing the system temperature $T$
across the Kosterliz-Thouless-Berezinskii point \cite{2005JSMTE..05..006D}.\\ 
\indent This scenario indicates that the problem of heat conduction in 2D systems is far from being completely explored and understood. In this perspective, it is important to investigate how anomalous heat transport changes in the transition between 2D- to quasi-1D and 1D systems.
Besides this motivation, it is also relevant to go beyond lattice models to assess 
the universality hypothesis in the more general contest of classical and quantum fluids
and even plasmas in low-dimensions.\\  
\indent In this paper we aim at exploring the above questions  in the context of a simple model for a two-dimensional plasma and to study its  
statistical properties as measured by the 
correlation functions of the fluctuations of the conserved fields. 
In particular, we will focus on a {\it one-component plasma}, hereafter OCP \cite{0022-3719-7-1-001,1980PhR....59....1B}. Such a model, despite its highly idealized nature, is suitable to treat a broad range of plasma regimes. For instance, OCP models have been applied to the study of relaxation in ultracold plasmas \cite{2007PhR...449...77K,2012PhRvL.109r5008B}, phase transitions in Coulomb crystals \cite{1987PhRvA..35.4743T,1990PhRvA..42.4972D,2008PhPl...15e5704B}, neutron-star crust crystallization \cite{1983ApJ...265L..83I,1995NCimA.108..431D,2012JPhCS.342a2005H}, cooling of magnetized plasmas \cite{1995AstL...21..702B,2015PhRvE..92f3105O}, degenerate inertial-fusion plasmas \cite{2006SPP....19.....P}, as well as charged colloids in solution \cite{1986PhRvL..57.2694K,2015PhRvE..91c2310R} and Yukawa liquids \cite{2008JPCM...20O3101D,2011PhRvE..84d6401M}. For an extensive review see  \cite{1999RvMP...71...87D} and references therein.\\
\indent The simulation studies are 
carried using the multiparticle collision algorithm (MPC) first introduced by Malevanets and Kapral \cite{1999JChPh.110.8605M,2004LNP...640..116M} and later widely employed for the simulation of the mesoscopic dynamics of polymers in solution, colloidal fluids and other complex fluids (e.g. see \cite{kapral08} and references therein). Such method is based on a mesh-dependent stochastic rule mixing particle velocities, constrained by the local conservation of kinetic energy, momentum and angular momentum. Application of the technique in plasma physics is,  at the best of our knowledge, new
\citep{2015PhRvE..92f2108D} and has its own interest as a promising 
tool to investigate a variety of problems, such as for example transport in complex  magnetized plasmas \cite{2015PhRvE..92f3105O,2016CoPP...56..246O}, discreteness effects in charged particle beams dynamics \cite{2000PhRvS...3c4202S,2004PhRvS...7a4202K}, as well as collision-driven transport of neutrals in fusion plasmas \cite{0029-5515-55-5-053025,2016NucFu..56l4002O}.\\
\indent The paper is structured as follows: in Section II we introduce the model and the main quantities of interests, in Section III we detail the numerical code (Multi-Particle 
Collision) used for the simulations, in Section IV we show the results for 2D systems and quasi-1D systems, with respect also to our previous results on the 1D version of the model, as well as the effect of a self-consistent electrostatic field. Finally, in Section V we summarize and point out the possible development  of this work. The Appendix  contains some  details on the implementation of conservation laws in the numerical code employed in this paper.  
\section{The Model}
We consider a OCP, namely a system  of $N_p$ charged particles of charge $q$ and mass $m$ (e.g. electrons),  embedded in a neutralizing and static homogeneous background (e.g. ions) with charge density $\varrho$. The state of a OCP is fully determined by a single macroscopic  quantity, the plasma coupling parameter, usually defined \cite{1999RvMP...71...87D,doi:10.1063/1.4900625} as the ratio of a typical nearest-neighbour interaction potential energy and mean thermal energy as 
\begin{equation}\label{plasmpar}
\Gamma\equiv\frac{\bar U}{k_BT}.
\end{equation} 
In the equation above, $k_B$ and $T$ are the Boltzmann constant and the plasma temperature (or the average particle kinetic energy $\langle K\rangle$ if the system is not in thermal equilibrium), respectively, while the form of the mean inter-particle Coulomb potential energy $\bar U$ depends on the dimensionality of the system and the screening of counter-charges \cite{2011PhPl...18f3701O}. Typically, it is assumed that
\begin{equation}\label{meanu}
\bar U=\frac{q^2}{4\pi\epsilon_0a},
\end{equation}
where $\epsilon_0$ is the permittivity of free space, and the Wigner-Seitz radius $a$ defines the average inter-particle distance as function of the number density $n$ as $(4\pi n/3)^{-1/3}$ in 3D, and $(2\pi n)^{-1/2}$ in 2D \cite{2008JPCM...20O3101D}.\\
\indent Hereby we consider a two-dimensional globally homogeneous neutral OCP, for which the typical interaction range is given by the 2D Debye length 
\begin{equation}\label{debye}
\lambda_D = \sqrt{\epsilon_0k_BTa/q^2n}.
\end{equation}
\indent It remains to introduce at this point the two principal time scales of the system, $t_{\rm dyn}$ and $t_{\rm coll}$, associated to the collective modes (e.g. the so-called Langmuir waves \cite{1992wapl.book.....S}), and to the collisionality of the system, respectively. In a two dimensional OCP the dynamical time $t_{\rm dyn}$ is related to the 2D plasma frequency $\Omega_P$ \cite{2008JPCM...20O3101D,2009PhRvE..79b6401D} by
\begin{equation}\label{plasmafreq}
t_{\rm dyn}=4\pi/\Omega_P;\quad\Omega_P=\sqrt{nq^2/2\epsilon_0am},
\end{equation}
while the collision time $t_{\rm coll}$ is the inverse of the collision frequency \cite{1965pfig.book.....S} and reads
\begin{equation}\label{collfreq}
t_{\rm coll}=1/\Omega_{\rm coll};\quad\Omega_{\rm coll}=\frac{nq^4\ln\Lambda}{2\pi a\epsilon_0^2m^{1/2}(k_BT)^{3/2}}.
\end{equation}
The expression for $\Omega_{\rm coll}$ has been rescaled ad hoc in order to account for the fact that the system is defined in 2D and $n$ has the meaning of a surface number density. In the equation above, the argument of the Coulomb logarithm $\ln\Lambda$ is somewhat arbitrary, we take here $\Lambda=\lambda_D/a_{\rm min}$, where the typical minimum inter-particle distance is usually $a_{\rm min}\approx a/10$ for our choice of parameters.\\   
\indent As we are primarily interested in the collision-driven energy transport, throughout this work we will consider only non-degenerate regimes for which $a<\lambda_D$, excluding for example ultra correlated plasmas (i.e. $\Gamma>200$) for which $a$ excedes $\lambda_D$, as well as extremely collisionless systems where $t_{\rm coll}\gg t_{\rm dyn}$ ($\Omega_{\rm coll}\ll\Omega_P$).\\
\indent In order to study the transport properties of the OCP, we measure the thermal conductivity $\kappa$ making use of the Green-Kubo formula
\begin{equation}\label{greenkubo}
\kappa=\frac{D}{k_BT^2N}\int_0^\infty \langle \mathbf{J}_\mathcal{E}(t) \mathbf{J}_\mathcal{E}(0) \rangle_{\rm eq} {\rm d}t,
\end{equation} 
where $D$ is a dimensional constant and $\langle \mathbf{J}_\mathcal{E}(t) \mathbf{J}_\mathcal{E}(0) \rangle_{\rm eq}$ is the equilibrium time--correlation function of the energy current
\begin{equation}\label{jene}
\mathbf{J}_\mathcal{E}(t)=\sum_{j=1}^{N_p}\mathcal{E}_j\mathbf{v}_{j}.
\end{equation}
For charged systems the particle energy $\mathcal{E}_j$ is given by
\begin{equation}
\mathcal{E}_j=\frac{m\mathbf{v}_j^2}{2}+q\Phi(\mathbf{r}_j),
\end{equation}
wherein $\Phi(\mathbf{r})$ is the electrostatic potential due to the charge distribution and/or, eventually, an external contribution. In the formulae above, $\mathbf{r}_j$ and $\mathbf{v}_j$ are particles positions and velocities.\\
\indent In 1D systems, where typically $\kappa\propto N^{\gamma}$, 
an effective way for obtaining the exponent $\gamma$ amounts to estimate the low frequency behavior of  \cite{2003PhR...377....1L,DHARREV} 
\begin{equation}
C_\mathcal{E}(\omega)=\langle|\hat \mathbf{J}_{\mathcal{E}}(\omega)|^2\rangle \sim 
\omega^{-\gamma};\quad {\rm for}\quad \omega\to 0,
\end{equation}
i.e. the Fourier transform of $\langle \mathbf{J}_\mathcal{E}(t) \mathbf{J}_\mathcal{E}(0) \rangle_{\rm eq}$.\\
\indent For 2D systems, instead, the logarithmic divergence of $\kappa$ with the size $N$ amounts to a $t^{-1}$ decay of the correlations which
is equivalent to 
\begin{equation}\label{omegalog}
C_\mathcal{E}(\omega)\sim \left[\alpha -\beta\log(\omega)\right];\quad {\rm for}\quad \omega\to 0, 
\end{equation}
where $\alpha$ and $\beta$ are two positive constants (see e.g. Ref. \cite{2005JSMTE..05..006D}, and references therein).\\
\indent In order to provide a complete description of the transport process of the model we analyze also the charge density current correlator $C_\rho(\omega)$, defined in the same fashion as $C_\mathcal{E}(\omega)$. The spatial density of a system of discrete charges $q$ in a homogeneous neutralizing background $\varrho$ is defined as
\begin{equation}\label{rhodisc}
\rho(\mathbf{r})=\varrho+\sum_{j=1}^{N_p}\left[q\delta(\mathbf{r}-\mathbf{r}_j)\right],
\end{equation}
so that the charge current $\mathbf{J}_\rho$ reads
\begin{equation}\label{jrho}
\mathbf{J}_\rho(t)=\sum_{j=1}^{N_p}\left[\varrho+q\delta(\mathbf{r}-\mathbf{r}_j)\right]\mathbf{v}_{j}.
\end{equation}
As we are going to discuss in Sec IV, a special importance for our analysis is played also by the density dynamical structure factor $S_{\rho}(\mathbf{k},\omega)$, containing information on the inter-particle correlations and their time evolution. This quantity is constructed in our numerical simulations as follows: first of all, we introduce the spatial Fourier transform of the density at a given time  $t$, that reads according to the definition of $\rho(\mathbf{r})$ given in Eq. (\ref{rhodisc}) (see also \cite{2015CoPP...55..421K}), as
\begin{equation}\label{pointrho}
\hat \rho(\mathbf{k},t) =\varrho\delta(\mathbf{k})+\frac{1}{N_p}\sum_{j=1}^{N_p}q\exp\left[{{\rm i}2\pi\mathbf{k}\cdot\mathbf{r}_{j}(t)}\right],
\end{equation}
where the first term arises from the definition of Fourier transform of a constant. We then take the temporal discrete Fourier transform of $\hat \rho(\mathbf{k},t)$ at fixed wave number $\mathbf{k}$ that yields
\begin{equation}\label{rhokomega}
\hat \rho(\mathbf{k},\omega) = \frac{1}{N_t}\sum_{l=1}^{N_t} \hat\rho(\mathbf{k},t_l)\left[\cos(-\frac{2\pi t_l\omega}{N_t})+{\rm i}\sin(-\frac{2\pi t_l\omega}{N_t})\right],
\end{equation}
where $N_t$ is the total number of equally sized time-steps $\Delta t$ performed by the simulation, so that $t_l = l \,\Delta t$. Finally, by taking the modulus square of $\hat\rho(\mathbf{k},\omega)$ we obtain  
\begin{equation}\label{somega}
S_{\rho}(\mathbf{k},\omega)=\langle|\hat \rho(\mathbf{k},\omega)|^2\rangle_{\rm eq}.
\end{equation}
Note that, in our numerical implementation, the temporal Fourier transform of charge density appearing in Eq. (\ref{rhokomega}) is computed only for a small number of wave vectors $\mathbf{k}$, thus avoiding to increase dramatically the memory load. On the other hand, we are primarily interested to
analyze the hydrodynamic limit of the model, that corresponds to consider only  low-$\mathbf{k}$ modes. Note also that, instead of evaluating $\hat\rho(\mathbf{k},t)$ as in Eq. (\ref{rhodisc}), one could in principle coarse grain the density on a mesh (cfr. Eq. (\ref{rhocont}) in the following section) and then take its time transform.
\section{The numerical code}
At variance with the pioneering numerical studies on the OCP based on direct molecular dynamics \cite{1977PhLA...63..301B,PhysRevA.18.2345,1979PThPh..62..883T}, and more recent numerical work involving particle-particle-particle mesh codes (P$^3$M, see \cite{1980CoPhC..19..215E}) \cite{2003PhRvL..90v6804D,2008JPCM...20O3101D}, in this work we adopt a novel computational approach, effectively splitting the Coulomb interaction in its short- and long-range contributions, treating them with a hybrid multiparticle-collision (MPC)--particle-in-cell (PIC) code.\\
\indent As in standard mesh-based computational schemes, the spatial domain of the simulation is coarse-grained into equal cells of size $\Delta s$. Inside each cell, Coulomb scatterings among particles are resolved {\it stochastically} by mixing in a collision step the particles velocities, so that their total momentum, kinetic energy and angular momentum are conserved; while during the ''streaming" step, the same are updated along with the associated position under the effect of the self-consistent electromagnetic field, computed on the grid with the usual PIC or particle-mesh technique \cite{1981csup.book.....H}. 
\subsection{The multiparticle collision scheme}
The MPC codes nowadays used in numerical complex fluid dynamics rely on different velocity exchange rules (see e.g. Ref. \cite{2009acsa.book....1G} for an extensive review). Here, we briefly review the general implementation of the widely used {\it stochastic rotation dynamics} (hereafter SRD).\\
\indent Let us consider a system of $N_p$ equal particles partitioned into $N_c$ equal volume cells in Cartesian coordinates. The particles move in continuum 2D space with momentum $\mathbf{p}_j=m\mathbf{v}_j$, either freely or under the effect of an external and/or self-consistent force field. 
In order to perform a collision step in the $i$--th cell one has to compute first its center of mass velocity  
\begin{equation}\label{mtotptot}
\mathbf{u}_i=\frac{1}{M_i}\sum_{j=1}^{N_i}\mathbf{p}_j;\quad M_i=\sum_{j=1}^{N_i} m=mN_i,
\end{equation}
where $N_i$ is the number of particles in the cell. The collision amounts to a rotation $\hat\mathbf{R}$ of an angle $\pm\varphi_i$ with probability one-half
of the  {\it relative} velocities $ \delta\mathbf{v}_j=\mathbf{v}_j-\mathbf{u}_i$, namely
\begin{equation}\label{rotation}
\mathbf{v}_{j}^\prime=\mathbf{u}_i+\hat\mathbf{R}_i\cdot\delta\mathbf{v}_{j}.
\end{equation}  
Such a rotation guarantees the conservation of the total momentum
and kinetic energy in the cell: 
\begin{eqnarray}\label{sist}
\mathbf{P}_i&=&\sum_{j=1}^{N_i} m\mathbf{v}_j=\sum_{j=1}^{N_i} m\mathbf{v}_j^\prime,
\end{eqnarray}
and
\begin{eqnarray}\label{sist1}
K_i&=&\frac{1}{2}\sum_{j=1}^{N_i} m\mathbf{v}_j^2=\frac{1}{2}\sum_{j=1}^{N_i} m\mathbf{v}_j^{\prime 2}.
\end{eqnarray}
However, with such a choice of the velocity rotation protocol, the total angular momentum $L_i$ in the cell is not conserved  \cite{doi:10.1021/jp046040x}.\\
\indent Several MPC algorithms that account for the angular momentum conservation do exist \cite{2008PhRvE..78a6706N,PhysRevE.92.013301,doi:10.1021/jp046040x}. In
this paper we impose also this conservation law by adopting the so-called deterministic rotation scheme (DR, originally introduced in \cite{tesiryder}, see also \cite{PhysRevE.78.016706,2009acsa.book....1G}) that applies only to 2D systems. In practice, the cell-dependent rotation angle $\varphi_i$ that defines $\hat\mathbf{R}_i$ in equation (\ref{rotation}) is evaluated deterministically from the relation
\begin{equation}\label{sincosphi}
\sin\varphi_i=-\frac{2a_ib_i}{a_i^2+b_i^2};\quad  \cos\varphi_i=\frac{a_i^2-b_i^2}{a_i^2+b_i^2},
\end{equation}
where the coefficients $a_i$ and $b_i$ are given as functions of particles' positions and velocities by
\begin{equation}\label{adef}
a_i=\sum_{j=1}^{N_i}\mathbf{r}_j \wedge (\mathbf{v}_j-\mathbf{u}_i),
\end{equation}
and
\begin{equation}\label{bdef}
b_i=\sum_{j=1}^{N_i}\mathbf{r}_j\cdot(\mathbf{v}_j-\mathbf{u}_i),
\end{equation}
where $\wedge$ denotes the external product in two dimensions. With such a choice of $\varphi_i$ the angular momentum conservation in cell $i$ reads
\begin{eqnarray}\label{sist2}
{L}_i&=&\sum_{j=1}^{N_i} m(\mathbf{r}_j \wedge \mathbf{v}_j)=\sum_{j=1}^{N_i} m(\mathbf{r}_j \wedge  \mathbf{v}_j^\prime).
\end{eqnarray}
The proof of the angular momentum conservation under a DR move is reported in Appendix, along with the proof of kinetic energy and linear momentum under the more general SRD scheme.\\
\indent Note that, since we are considering point-like particles, the contribution of an internal degree of freedom associated to particle size (i.e. a classical spin) does not enter the definition of $L$ and its local conservation under MPC dynamics.  However, due to the imposed {\it periodic} boundary conditions (PBC) in our simulation set-up, $L$ is not
globally conserved.  In practice, the angular momentum $l_i$ of a particle $i$ of mass $m$ with velocity $\mathbf{v}_i=(v_{xi};v_{yi})$ changes as the latter crosses an edge of the simulation domain (e.g., $x_{\rm max};y$), and it is re-injected at the opposite one ($x_{\rm min};y$), i.e.:
\begin{equation}\label{noangularmom}
l_i=m(x_{\rm max}v_{yi}-y_iv_{xi})\neq m(x_{\rm min}v_{yi}-y_iv_{xi}).
\end{equation}
It is important to remark at this stage, that in order to correctly reproduce the hydrodynamics of the system, the conservation rules should indeed be {\it local} (i.e. at the cell level in our case), as proved in \cite{2011PhRvL.106u0601B}, and therefore the violation of the global conservation of $L_{\rm tot}$ due to the choice of PBC is irrelevant. In fact, in our simulations we always start with null total angular momentum and the fluctuations due to the boundary effect average to zero.\\
\indent Moreover, note also that, with the implementations of the MPC method described here, the Galilean invariance of the particle equations of  motion is violated. To avoid this complication, before the collision step all particles of the simulation are shifted by the same vector $\mathbf{d}$ with components  $d_x, d_y$, chosen from a uniform distribution between $\Delta s/2$ and $-\Delta s/2$, where $\Delta s$ is the cell size. After the collision, the particles are shifted back of $-\mathbf{d}$ to their original position. It has been proved, that if the system mean path $\lambda_{\rm coll}>\Delta s/2$, the violation of the Galilean invariance is negligible \cite{2001PhRvE..63b0201I,2003PhRvE..67f6705I}.\\
\indent Up to now, we reviewed the SRD and DR in the standard fluid case. In a series of papers on the anomalous diffusion and heat transfer in 1D one-component plasmas \cite{2010JPhCS.260a2005B,2013PhRvE..87b3102B,2015PhRvE..92f2108D}, we have adapted a reduced version of the MPC technique to treat a fluid of particles interacting via effective Coulomb forces, by conditioning the velocity exchange to an interaction probability $\mathcal{P}_i$, which depends on the values of the plasma parameters in the cell.\\    
\indent In this work, we proceed in the same fashion introducing for each cell the {\it local} plasma coupling parameter (cfr. Eqs.(\ref{plasmpar}-\ref{meanu})) $\bar\Gamma_i = \bar U_i/\bar K_i$, where $\bar U_i$ and $\bar K_i=K_i/N_i$ are the mean interparticle potential energy and mean kinetic energy in cell $i$, respectively. In order to account for the logarithmic nature of the Coulomb interaction in two dimensions \cite{kellogg}, $\bar U_i$ is corrected by the multiplicative factor $-\log(a_i/\Delta s)$, where $a_i=(N_i/\Delta s^2)^{-1/2}$. Note that, in the range of parameters considered here, such quantity is always positive and of order 1.\\
\indent During the collision step, the multiparticle collision probability is evaluated as 
\begin{equation}\label{prob}
 \mathcal{P}_i=\frac{1}{1+\bar\Gamma_i^{-2}}.
\end{equation}
By sampling a random number $\mathcal{P}_i^*$ from a uniform distribution in the interval $[0,1]$, the rotation (i.e. the multi-particle collision) (\ref{rotation}) runs if $\mathcal{P}_i^*/\mathcal{P}_i\leq1$.\\
\indent Note that, the formulae above are written for a single-mass system. However, several generalizations of the MPC techinque to the case of multi-mass systems do exist (see e.g. Ref. \cite{kapral08}). Hereafter, we will only deal with single species systems, where all $m_j=m$.\\
\indent 
From a practical point of view this probabilistic interaction rule, inspired by heuristic arguments, is equivalent to adopt a distribution of the time between collision
events in each cell.
Translation invariance guarantees that this distribution is independent of cell $i$. We have also checked that in a wide range of parameters
this distribution is Poisson-like and its typical time scale depends on $\Gamma$.
\subsection{Computation of the self-consistent electrostatic field and tests}
In order to obtain a more complete picture of the transport properties of the system, we also study the contribution of its self-consistent electrostatic field $\mathbf{E}$, evaluated with the standard particle-mesh technique with a Fourier space-based Poisson-solver \cite{1981csup.book.....H}. In the numerical calculations presented in this paper, we consider 2D systems in a rectangular simulation box with periodic boundary conditions, partitioned in $N_c=N_x\times N_y$ equal square cells of size $\Delta s$. In each cell the charge density $\rho_{i,j}$ is given by
\begin{equation}\label{rhocont}
\rho_{i,j}=\varrho_{i,j}+\frac{1}{\Delta s^2}\sum_{k=1}^{N_{i,j}}q_k,
\end{equation} 
where $N_{i,j}$ is the number of particles in the cell while $q_k$ are their charges. For the sake of simplicity we assume that the fixed neutralizing background 
density is everywhere constant, i.e. $\varrho_{i,j} = \varrho$. 
In practice, the electrostatic field can be evaluated by the standard equation $\mathbf{E}(\mathbf{r}) =-\nabla\Phi(\mathbf{r})$, where the electrostatic potential 
$\Phi(\mathbf{r})$ is related to the charge density by the Poisson equation $\Delta\Phi(\mathbf{r})=\rho(\mathbf{r})/\epsilon_0$, which is easier to be solved in
Fourier space, see e.g. \cite{2000NewA....5..305F}, and references therein.\\
\begin{figure}
\begin{center}
\includegraphics[width=0.8\columnwidth]{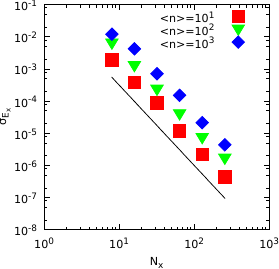}
\caption{Fluctuations $\sigma_{\mathbf{E}_x}$ of the $x$-component of the electrostatic field as function of the cell number $N_x$ along $x$, for different values of the particle number density $n$. The solid line marks the fitting power-law trend obtained by the data points $\sigma_{\mathbf{E}_x}\propto N_x^{5/2}$.}
\label{fluctuation}
\end{center}
\end{figure}
\indent In our code, the Fourier transforms are computed with the publicly available {\sc fftw} package \cite{2012ascl.soft01015F}. 
Eventually, when $\Phi(\mathbf{r})$ is obtained the electric field is evaluated at each particle position by standard two-dimensional interpolation procedures \cite{2002nrca.book.....P}.\\
\indent The particles equations of motion under the effect of the electric field $\mathbf{E}$ are integrated in our {\sc fortran90} code (see also \cite{2014arXiv1408.3857D} for further details), with the standard second order {\it leapfrog} scheme, widely used in molecular dynamics simulations \cite{doi:10.1080/08927029108022142,1995PhyS...51...29C}. For all simulations presented here we use a {\it bona fide} fixed timestep $\Delta t=0.05t_{\rm dyn}$ ensuring energy conservation up to 1 part in $10^{-12}$ when using double precision, while still allowing for acceptable computational times on a single core of an i5 HP\textregistered~ machine running {\sc linux}.\\
\indent In the present paper we investigate only periodic systems with global charge neutrality, characterized by {\it equilibrium} phase-space distribution function
\begin{equation}\label{df}
f(\mathbf{r},\mathbf{v})=\frac{\mathcal{C}n}{2\pi mk_BT}\exp(-m\mathbf{v}^2/2k_BT),
\end{equation}
where $n$ is the (spatially constant) number density and $\mathcal{C}$ is a normalization factor so that the integral of $f$ over the simulation domain equals 1.\\
\indent Note that for this class of initial conditions, the  average self-consistent electrostatic field is  zero,
because the counter background charge screens the long-range tail of the Coulomb interaction. 
However, spatiotemporal fluctuations of the field $\mathbf{E}(\mathbf{r}_i)$ persist. We performed test simulations of globally neutral equilibrium systems for different values $n$ and different combinations of system size and grid resolution. We found that, for $N_c=N_x\times N_y\geq 50$, the electrostatic field averaged over the particle positions is actually zero, independently on the systems size. For fixed $n$ and fixed cell size $\Delta s$, the amplitude of its fluctuations $\sigma_{\mathbf{E}}$ decrease with the systems size as a power-law as shown in Fig. \ref{fluctuation} for the $x$-component of $\mathbf{E}$.\\
\indent 
In this paper, we want also to address the question if the presence of such fluctuations of $\mathbf{E}$ have an influence on the hydrodynamic behavior of 2D neutral plasmas.
In the following Section, we report two sets of numerical experiments for fixed plasma parameters. In the first case we impose $\Phi=0$, so that the conserved total energy is reduced to the kinetic term only, and particles move freely between collisions. In the second case $\Phi$ is computed from the instantaneous distribution of particles whose
dynamics depends also on the fluctuating field $\mathbf{E}$. Despite the amplitude of fluctuations in the explored regimes is quite small, the presence of the fluctuating field could yield some changes in 
the hydrodynamic behavior of the system. In fact, as discussed 
in the following section, it does not affect significantly the form of the structure factors, but, this notwithstanding, 
the low--frequency component of the energy current frequency spectrum exhibits some difference, that
can be attributed to finite size effects.\\
\indent In the present work we do not investigate regimes where the self-consistent electric field is
large, as it happens for sensible charge unbalance or in the presence of an external potential. 
These cases will be analyzed in a forthcoming publication.\\
\indent All numerical simulations presented in this paper have been carried out making use of units such that $k_B=\epsilon_0=m=q^2=1$, while the normalization of distances is fixed so that the cell length $\Delta s=\lambda_D=1$. With such a choice, the numerical model has only two control parameters, i.e. the temperature $T$ and the average number density $n$ that combined together yield $\Gamma$.
\section{Results}\label{seciv}
\begin{figure}
\begin{center}
\includegraphics[width=0.8\columnwidth]{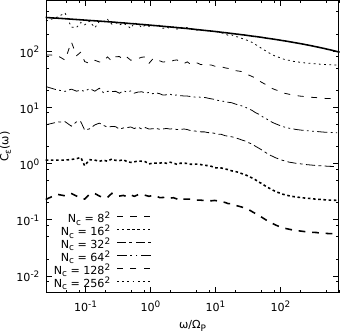}
\end{center}
\caption{Fourier spectra $C_\mathcal{E}$ of the energy current, for $\Gamma= 3$, and $N_c=8^2$, $16^2$, $32^2$, $64^2$, $128^2$, and $256^2$ (dashed lines). To guide the eye, the fitting function $f(\omega)\propto \alpha-\beta\log(\omega)$ is added to the figure (heavy solid line).}
\label{current2d}
\end{figure}
\subsection{Two dimensional systems}
In a first set of numerical simulations we study the behavior of the energy and density correlators of 2D OCP
for different systems sizes and values of $\Gamma$. The initial conditions are generated by sampling particles positions and velocities from the phase-space distribution (\ref{df}) for the chosen values of temperature $T$ and particle density $n$.
\begin{figure}
\begin{center}
\includegraphics[width=0.8\columnwidth]{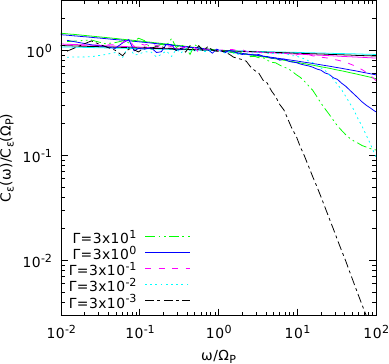}
\end{center}
\caption{Normalized Fourier spectra of the energy current $C_\mathcal{E}$ for 2D systems with different values of the plasma coupling parameter $10^{-3}<\Gamma< 10^{2}$ (dashed lines). The frequency is in units of $\Omega_P$ as given in Eq. (\ref{plasmafreq}), while the correlator scale is in units of the corresponding $C_\mathcal{E}(\Omega_P)$. For each case, the best-fit curve according to Eq. (\ref{omegalog}) is added (solid lines).}
\label{fits}
\end{figure}
The particles equations of motion have been integrated over a time scale $t_{\rm end}\approx 7000t_{\rm dyn}$.  Such a choice guarantees a good convergence to equilibrium over  the explored range of parameters.\\
\indent  The main result of our study is that the correlator of the energy current $C_\mathcal{E}(\omega)$ always shows a clear logarithmic behavior for low $\omega$,
as expected on the basis of general theoretical arguments (see e.g. \cite{2003PhR...377....1L}).\\
\indent In Fig.~ \ref{current2d} we show $C_\mathcal{E}(\omega)$ for an OCP with $\Gamma= 3$, for different system sizes ranging from $N_c=8^2$ up to $256^2$. All curves appear to be well fitted by Eq. (\ref{omegalog}) in the interval of frequencies $10^{-2}\leq\omega/\Omega_P\leq 10$.\\
\indent The robustness of this logarithmic scaling can be tested while varying $\Gamma$ (e.g. varying $T$ at fixed $n$ or, vice-versa fixing $T$ and varying $n$). In Fig.~\ref{fits} we report the normalized quantity
 $C_\mathcal{E}(\omega)/ C_\mathcal{E}(\Omega_p)$ versus $\omega/\Omega_p$ for fixed system size (in units of its $\lambda_D$) while varying $\Gamma$ over five orders of magnitude. We find that the logarithmic fit is maintained and is optimal for $\Gamma \approx 1$, which is at the
border between strong and weak coupling regimes. In addition, we have also checked that simulations with initial conditions characterized by different combinations of $T$ and $n$ yielding the same values of $\Gamma$ are associated with qualitatively similar results.
\begin{figure}
\begin{center}
\includegraphics[width=0.8\columnwidth]{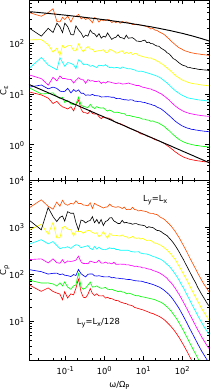}
\end{center}
\caption{Fourier spectra of the energy current $C_{\mathcal{E}}$ (top panel) and of the density current $C_\rho$ (bottom panel) as function of the frequency $\omega$ normalized to $\Omega_P$, for $\Gamma= 3$, and $L_x/L_y=128$, 64, 32, 16, 8, 4, 2, and 1. The curves are averaged over 100 independent realizations. The two heavy solid lines in the upper panel mark the predicted $\omega^{-1/3}$ and $\alpha-\beta\log\omega$ trends in pure 1D and 2D cases, respectively.}
\label{fcorrs}
\end{figure}
\subsection{Dimensional cross-over}
In the previous section we have checked  the expected logarithmic 
divergence (\ref{omegalog}) of the energy current correlator of the 2D OCP model. Here we investigate how such a behavior crosses over to
the power-law behavior predicted by the KPZ hydrodynamics when passing from 2D to quasi-1D systems. 
\\
\indent In the simulations reported hereafter we fix $\Gamma=3$ (i.e., moderately coupled particles) and
$N_x=256$, while $ 2 \leq N_y \leq 256$ (i.e., $ 1/128 \leq L_y/L_x \leq 1$). 
Notice that for the adopted value of $\Gamma$, $\Omega_{\rm coll}>\Omega_P$, so that the contribution of the fluctuating
\begin{figure}
\begin{center}
\includegraphics[width=0.9\columnwidth]{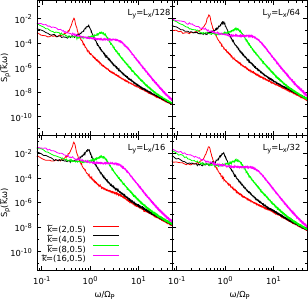}
\caption{Dynamical structure factor of density $S_{\rho}$ for $\tilde\mathbf{k} = (2,0.5)$, (4,0.5), (8,0.5), and (16,0.5), $N_x=256$ and $N_y=2$ (top left), 4 (top right), 8 (bottom right), and 16 (bottom left). In all cases $\Gamma=3$.}
\label{peaksunscaled}
\end{center}
\end{figure}
electrostatic field on the collisional dynamics is very small. For the sake of simplicity in these simulations we have
set $\mathbf{E} = 0$. In the following sub-section we shall analyze also the effects of a nonzero electrostatic
field.\\
\indent In Fig.~\ref{fcorrs} we show the Fourier spectra of the energy and density current correlators 
$C_\mathcal{E}(\omega)$
(upper panel) and $C_\rho(\omega)$ (lower panel)
for different values of $N_y$. For small values of $N_y$ $C_\mathcal{E}(\omega)$  exhibits a $\omega^{-1/3}$ slope
for small values of $\omega$, typical of 1D systems with three conservation laws, while the 
logarithmic singular behavior is recovered for sufficiently large value of $N_y$. In particular,
the crossover between these different scaling laws can be approximately identified for $N_y=8$ (i.e. $L_y=L_x/16$, see the third curve from below).\\
\indent For what concerns $C_\rho(\omega)$ for small values of $N_y$ we recover the same power--law
behavior observed in \cite{2015PhRvE..92f2108D} for a 1D OCP. 
When $N_y$ is increased the exponent of the power
law seems just to decrease. We conjecture that a logarithmic singularity could be recovered also
for $C_\rho(\omega)$ by simulating much larger systems, a check that is far beyond our computational
resources.\\
\indent Moreover, we have also computed the density structure factor $S_{\rho}(\mathbf{k},\omega)$, 
that has been 
used as a testbed to check  the validity of KPZ fluctuating hydrodynamics in 1D OCP 
(see Figs. 6-7 in Ref. \cite{2015PhRvE..92f2108D}).\\ 
\indent In Fig.~\ref{peaksunscaled} we show $S_{\rho}(\tilde\mathbf{k},\omega)$ for  
$N_y$ significantly smaller than $N_x= 256$. For each value of $N_y$ we report the data 
corresponding to four low values of the normalized wave number $\tilde{\mathbf{k}}=(2, 0.5)$, $(4,0.5)$, $(8,0.5)$, and $(16,0.5)$, that point out the hydrodynamic limit of the model. As  already
observed for 1D models (cfr. Figs. 1 and 6 in Ref. \cite{2015PhRvE..92f2108D}),  also these curves exhibit  a peak 
at $\omega_{\rm max}\propto c_s||\mathbf{k}||$ ($c_s$ is the sound velocity of the system) that sharpens for decreasing values of $\tilde k$ and $L_y$.\\
\begin{figure}
\begin{center}
\includegraphics[width=\columnwidth]{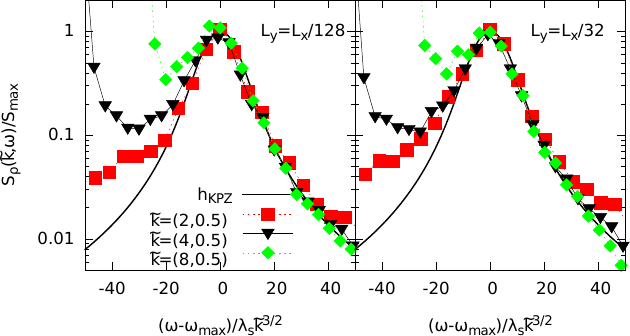}
\end{center}
\caption{Data collapse of the rescaled structure factor (points) to the KPZ scaling function $h_{\rm KPZ}$ (solid lines) for the modes corresponding to $\tilde\mathbf{k}=(2,0.5)$ (squares), (4,0.5) (downward triangles), and (8,0.5) (diamonds). The left panel refers to $N_y=2$ (i.e. $L_y=L_x/128$), and the right panel to $N_y=8$ (i.e. $L_y=L_x/32$).}
\label{scalingkpz}
\end{figure}
\indent The prediction of the  NFH theory \cite{2014JSP...154.1191S} for 1D systems is that 
the density correlation in the 
large-time and space scales should obey the dynamical scaling of the KPZ equation.
Accordingly, the structure factor $S(k,\omega)$ for small enough wave numbers $k$ and 
$\omega\approx\pm \omega_{\rm max}$, are expected to scale as
\begin{equation}\label{scaling32}
S_{\rho}({k},\omega)\sim h_{\rm KPZ}\left(\frac{\omega-\omega_{\rm max}}{\lambda_s k^{3/2}}\right),
\end{equation}
where $\lambda_s$ is a model dependent coefficient that can be evaluated in terms of equilibrium
correlators, and $h_{\rm KPZ}$ is the universal KPZ scaling function that is not known in terms of simple functions \citep{2014JSP...154.1191S}. Asymptotic and integral forms of Eq. \ref{scaling32} are given e.g. in \cite{Prahofer2004}.\\
\indent It becomes natural to ask whether (and to which extent) the peaks of $S_\rho(\tilde\mathbf{k},\omega)$ are fitted by the KPZ scaling function. In order to test 
the quality of the fit, we have rescaled the longitudinal component of $S_\rho(\tilde\mathbf{k},\omega)$ according to Eq. (\ref{scaling32}), for the cases presented in Fig.~\ref{peaksunscaled}. In Fig.~ \ref{scalingkpz} we show that the structure factors obtained for $N_y=2$ and 8 exhibit a good data collapse
onto the KPZ scaling function for $\tilde \mathbf{k} = (2,0.5)$, (4, 0.5) and (8, 0.5). This analysis indicates that the system 
maintains the same hydrodynamic features of a genuine 1D system.
Moreover, also in the cases reported in Fig.~\ref{scalingkpz} the data collapse is very poor for larger values
of  $\tilde \mathbf{k}$ (data not shown), because of the presence of the heat mode peak at low values of $\omega$ 
\cite{2015CoPP...55..421K}.\\ 
\begin{figure}
\begin{center}
\includegraphics[width=0.9\columnwidth]{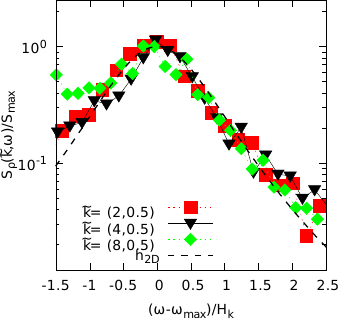}
\end{center}
\caption{Data collapse of the rescaled structure factor for a 2D system with $N_c=256\times 64$ (points) onto the scaling function $h_{2D}$ given by Eq. (\ref{stfr}) (dashed line), for the modes corresponding to $\tilde\mathbf{k}=(2,0.5)$ (squares), (4,0.5) (downward triangles), and (8,0.5) (diamonds).}
\label{scaling2d}
\end{figure}
\indent Conversely, we expect that approaching the 2D limit $N_y \sim N_x$ the data collapse on the KPZ scaling function will not
hold for small values of $\tilde\mathbf{k}$. As shown in Fig.~\ref{scaling2d}
for $N_y  = 64$ we still obtain for $\tilde \mathbf{k} = (2, 0.5)$, (4, 0.5) and (8, 0.5) a good data collapse of the structure factors, which
can be fitted empirically by a rational function 
\begin{equation}
\label{stfr}
S_{\rho}({k},\omega)\sim h_{2D}\equiv C\left[\left(\frac{\omega-\omega_{\rm max}}{H_k}\right)^\xi+1\right]^{-\zeta},
\end{equation}
where $H_k$ is the full-width-at-half-maximum of the sound peak, $C$ is scale factor depending on the normalization choice of $\omega$, and the numerical estimates of the
exponents yield $\xi \approx 2$ and $\zeta\approx 2$, seemingly independently on $\tilde \mathbf{k}$. We note that, independently on the normalization choice for $\omega$ and $k$, $H_k\sim k^{1.8}$.\\
\begin{figure}
\begin{center}
\includegraphics[width=0.8\columnwidth]{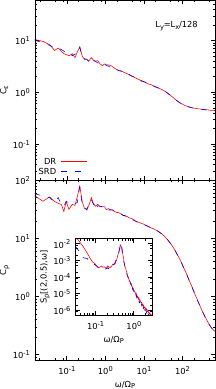}
\end{center}
\caption{Fourier spectra of the energy current $C_{\mathcal{E}}$ (top panel), and of the density current $C_\rho$ (bottom panel) for $\Gamma= 3$, $N_c=256\times 2$. The solid lines refer to simulations done with angular momentum conserving DR scheme, while the dashed lines refer to simulations using the standard SRD. The small inset in the bottom panel shows the dynamical structure factor of the density for $\tilde\mathbf{k}=(2,0.5)$. Note how the position and height of the sound peak is not altered.}
\label{noangular}
\end{figure}
\indent Having established the robustness of the NFH predictions even for non-perfectly 1D systems, as well as the expected universal behavior of the energy correlators for 2D systems, it is interesting to observe what happens if the local conservation of the angular momentum $L$ is violated (i.e. the number of local conservation rules is reduced).\\
\indent We repeated the numerical experiments described up to now with the same set-up, but using  the SRD rule to treat the Coulomb collisions. Surprisingly, no evidence of a somewhat different behavior of both $S_\rho$ and $C_\mathcal{E}$ is found, independently on the system size and/or transversal to longitudinal size ratio. As an example, in Fig.~\ref{noangular} we show the energy (top panel) and density current (bottom panel) correlators as function of $\omega$ for the case of quasi-1D system with $N_y=2$ (i.e. $L_y=L_x/128$) for simulations using DR and SRD protocols (i.e. with and without local conservation of $L$). The curves do not differ significantly bearing the same $\omega^{-1/3}$ slope at low frequency. In the small inset we also show the density structure factor for $\tilde\mathbf{k}=(2,0.5)$. Also for this quantity no appreciable difference is found, with the sound peak non appearing to change its position and height, thus implying the persistence of the data collapse to the KPZ scaling function $h_{\rm KPZ}$ (cfr. left panel of Fig. \ref{scalingkpz}).
\begin{figure}
\begin{center}
\includegraphics[width=0.8\columnwidth]{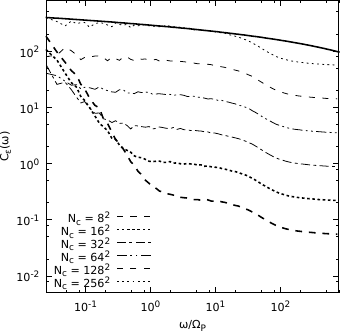}
\end{center}
\caption{Fourier spectra $C_\mathcal{E}$ of the energy current, for systems with active self consistent $\mathbf{E}$, $\Gamma= 3$, and $N_c=8^2$, $16^2$, $32^2$, $64^2$, $128^2$, and $256^2$ (dashed lines). To guide the eye, the fitting function $f(\omega)\propto \alpha-\beta\log(\omega)$ is added to the figure (heavy solid line).}
\label{current2dmf}
\end{figure}
\subsection{Effect of the self-consistent $\mathbf{E}$} 
It remains to determine the effect of the fluctuations of a globally null electrostatic field on the hydrodynamics of 2D and quasi-1D OCP. We have 
performed a set of numerical simulations by adding  the self-consistent electrostatic field $\mathbf{E}$, while maintaining the same values for all the other physical parameters.\\
\indent As anticipated in Section III for the typical system sizes considered here the fluctuations $\delta\mathbf{E}$ are of the order of $10^{-6}$. In Fig.~ \ref{current2dmf} 
we show the energy current correlator for the same 2D systems of Fig. \ref{current2d}. We observe that only small systems (i.e. $N_c\leq32^2$) are significantly  affected by the  presence of the fluctuating electric field. In particular, it corresponds to the presence of a noisy-like spectrum, i.e. $C_\mathcal{E}(\omega) \sim \omega^{-2}$, for $\omega<\Omega_P$, showing that the incoherent fluctuations of  $\mathbf{E}$ are typically slower that the period associated to the fundamental plasma frequency. This confirms that
the fluctuating self-consistent electric field does not affect the collective behavior of large enough systems \\
\indent We have also checked (data not reported) that the crossover from the $\omega^{-1/3}$ power-law divergence to the $\alpha-\beta\log(\omega)$ one,  
observed for $C_\mathcal{E}(\omega)$  when passing from 2D to quasi-1D systems, is unaffected by the presence of $\mathbf{E}$.\\
\begin{figure}
\begin{center}
\includegraphics[width=0.8\columnwidth]{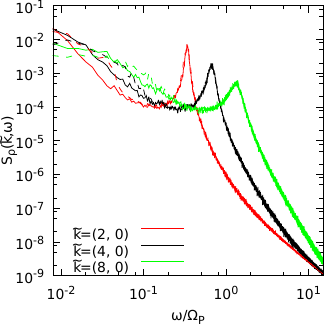}
\end{center}
\caption{Dynamical structure factor of density for a quasi-1D system with $L_y=L_x/128$ and $\Gamma= 3$, for $\tilde\mathbf{k}= (2,0)$, (4,0) and (8,0). Dashed lines refer to the simulations with activated $\mathbf{E}$, while the solid lines to those with $\mathbf{E}$ set to 0.}
\label{mfnomf}
\end{figure}
\indent Also the relevant features of the density structure factor $S_\rho(\tilde\mathbf{k},\omega)$ do not change with the presence of 
 $\mathbf{E}$.  In Fig. \ref{mfnomf} we  show this quantity  for three values of the normalized wave vector $\tilde\mathbf{k}=(2,0)$, (4,0) and (8,0), 
comparing the results of simulations with the zero-field case: we observe deviations only for small values of $\omega/\Omega_P$.
In summary, all the result discussed here and at the end of subsection B point out that  the hydrodynamics of (quasi)-1D systems  is robust with respect to 
the addition of  the angular momentum  conservation law as well as to the presence of a ``symmetry breaking" mechanism associated to the 
self-consistent fluctuating electric field.
\section{Summary and conclusions}
In this paper we have investigated the dynamical structure factors of density and the energy correlators of the One Component Plasma model over a few decades in the coupling parameter $\Gamma$. The main results are listed hereafter.\\ 
\indent When moving from a quasi one-dimensional setup to a two-dimensional one, we observe a cross over of $C_\mathcal{E}(\omega)$
from a power--law to a logarithmic divergence at small values of $\omega$. Such a hydrodynamic behavior indicates that the thermal 
conductivity $\kappa$ diverges with the system size as $\kappa \sim N^{1/3}$ for 1D systems and as   $\kappa \sim \log{N}$  for 2D systems.\\
\indent This picture is confirmed also by the form of the structure factors $S_{\rho}(\mathbf{k},\omega)$ that are fitted by the
KPZ scaling function for quasi 1D systems and by  a suitable rational function (\ref{stfr})  for 2D systems.\\
\indent This numerical results seem to suggest that also in the 2D case it exists a scaling
function, that should stem from a suitable hydrodynamic theory. 
Working out such a theoretical approach to the hydrodynamics of 2D OCP goes beyond
the aims of this paper and will be open to future investigations.\\
\indent When the angular momentum conservation law is removed, we do not observe any significant change of the previous results, apart
the presence of finite size effects for small values of $\omega$. This indicates that modes associated with angular momentum conservation
have no practical influence on the hydrodynamics of the model in 1D as well as in 2D systems.\\
\indent The addition of a self-consistent electrostatic field $\mathbf{E}$ to the plasma dynamics also reveals immaterial to the 
hydrodynamic properties of the model, at least for small amplitude fluctuations of the field. In fact, in the explored parameter space
the average value of $\mathbf{E}$ vanishes (neutral plasma), while the amplitude of its fluctuations $\sigma_{\mathbf{E}}$ is typically of $\mathcal{O}(10^{-6})$
and further decreases with the system size.
We cannot exclude that for larger amplitudes the overall scenario might change significantly. In addition, we point out that, substituting $\mathbf{E}$ with an opportunely tuned zero-average stochastic field $\mathbf{E}_s$, with fluctuations with amplitude of the same order of $\sigma_\mathbf{E}$, will not lead to the same conclusions. The reason of this being that when $\mathbf{E}$ is evaluated self-consistently the dynamics of charge density fluctuations $\delta\rho$ and field fluctuations $\delta\mathbf{E}$ are linked by Maxwell equations, while on the other hand, density fluctuations obviously can not have any effect on an externally imposed field.\\
\indent The natural follow-up of this work is the extension of our investigation to the case of three dimensional systems where a source of anisotropy is introduced, such as for instance, an axial magnetic field $\mathbf{B}_z$ is turned on. In this case energy transport is expected to work differently along and perpendicularly to the direction of $\mathbf{B}_z$. Moreover, as mentioned previously, hybrid MPC-PIC schemes seem to be promising for the modelization of plasma regimes in which the interplay between collisions and macroscopic electromagnetic fields is strong, such as for example the formation of run-away electrons in tokamak plasmas \cite{2016arXiv161003249S,2016NucFu..56k2009S}. A paper exploring this line is currently in preparation.
\section*{Acknowledgements}
We thank V. Popkov, P. Ghendrih and F. Piazza for the stimulating discussions at an early stage of this project. This work is part of the project CEA-01 ``ESKAPE" EUROfusion Enabling Research work programme 2017. G.C. would like to acknowledge the support from A*MIDEX project (Nr. ANR-11-IDEX-0001-02) funded by the ``Investissements d'Avenir" French Government program, managed by the French National Research Agency (ANR). P.F.D.C. acknowledges the support by the INFN initiative DYNSYSMATH 2016.
\section*{APPENDIX: Local conservation rules in SRD and DR}\label{proofconservation}
We prove here the local conservation rules in SRD and DR schemes. For reasons of simplicity we set particle masses $m=1$ so that in Eq. (\ref{mtotptot}) $M_i=N_i$.\\
\indent In order to check the conservation of linear momentum $\mathbf{P}_i$ within the $i-$th cell under the SRD rule, let us substitute the definition of particles relative velocities (\ref{rotation}) in the r.h.s. of Eq (\ref{sist}). Making use of the definition of cell center of mass velocity $\mathbf{u}_i=(\sum_j\mathbf{v}_j)/N_i$, one has
\begin{eqnarray}
\sum_{j=1}^{N_i}(\mathbf{u}_i+\hat\mathbf{R}_i\mathbf{v}_j-\hat\mathbf{R}_i\mathbf{u}_i)&=&N_i\mathbf{u}_i+\hat\mathbf{R}_i\sum_{j=1}^{N_i}\mathbf{v}_j-
\hat\mathbf{R}_i\mathbf{u}_i=\nonumber\\
=\sum_{j=1}^{N_i}\mathbf{v}_j+\hat\mathbf{R}_iN_i\sum_{j=1}^{N_i}\mathbf{v}_j
&-&\hat\mathbf{R}_iN_i\sum_{j=1}^{N_i}\mathbf{v}_j=\sum_{j=1}^{N_i}\mathbf{v}_j,
\end{eqnarray}
that proves the equality.\\
\indent The conservation of (twice) the kinetic energy $2K_i$ procedes in the same fashion by substituting the definition of $\mathbf{v}^{\prime}_i$ in Eq. (\ref{sist1}) so that it now reads
\begin{eqnarray}
\sum_{j=1}^{N_i}(\mathbf{u}_i+\hat\mathbf{R}_i\mathbf{v}_j-\hat\mathbf{R}_i\mathbf{u}_i)\cdot
(\mathbf{u}_i+\hat\mathbf{R}_i\mathbf{v}_j-\hat\mathbf{R}_i\mathbf{u}_i)=\nonumber\\
=N_i\mathbf{u}_i^2+\sum_{j=1}^{N_i}\left(\hat\mathbf{R}_i\mathbf{v}_j\right)^2+\sum_{j=1}^{N_i}\left(\hat\mathbf{R}_i\mathbf{u}_i\right)^2+\nonumber\\
+2\mathbf{u}_i\cdot\sum_{j=1}^{N_i}\hat\mathbf{R}_i\mathbf{v}_j-\sum_{j=1}^{N_i}2(\mathbf{u}_i\cdot\hat\mathbf{R}_i\mathbf{u}_i)+\nonumber\\
-2\sum_{j=1}^{N_i}(\hat\mathbf{R}_i\mathbf{v}_j)\cdot\hat\mathbf{R}_i\mathbf{u}_i=\nonumber\\
=2N_i\mathbf{u}_i^2+\sum_{j=1}^{N_i}\mathbf{v}_j^2-2N_i\hat\mathbf{R}_i\mathbf{u}_i\cdot\hat\mathbf{R}_i\mathbf{u}_i=\nonumber\\
=2N_i\mathbf{u}_i+\sum_{j=1}^{N_i}\mathbf{v}_j^2-2N_i\mathbf{u}_i=\sum_{j=1}^{N_i}\mathbf{v}_j^2,
\end{eqnarray}
where we have used the relation $\hat\mathbf{R}_i\mathbf{u}_i\cdot\hat\mathbf{R}_i\mathbf{u}_i=\mathbf{u}_i^2$.\\
\indent So far, the conservation of momentum and kinetic energy in the cell is verified for every rotation matrix $\hat\mathbf{R}$. In order to conserve angular momentum, the DR scheme poses a constraint on the choice of the rotation angle $\varphi_i$. Let us assume that the rotation matrix $\hat\mathbf{R}_{\varphi_i}$ verifies identity (\ref{sist2}), therefore
\begin{eqnarray}\label{angularconserv}
\sum_{j=1}^{N_i}\mathbf{r}_j\wedge\left[\mathbf{u}_i-\hat\mathbf{R}_{\varphi_i}(\mathbf{v}_j-\mathbf{u}_i)\right]=\nonumber\\
=N_i\mathbf{r}_i^c\wedge\mathbf{u}_i+\sum_{j=1}^{N_i}\mathbf{r}_j\wedge\hat\mathbf{R}_{\varphi_i}(\mathbf{v}_j-\mathbf{u}_i)=\nonumber\\
=N_i\mathbf{r}_i^c\wedge\mathbf{u}_i+\sum_{j=1}^{N_i}\mathbf{r}_j\wedge\hat\mathbf{R}_{\varphi_i}\mathbf{v}^c_j.
\end{eqnarray}
In the equality above $\mathbf{r}_i^c=(\sum_j \mathbf{r}_j)/N_i$ is the position of the centre of mass of cell $i$ and $\mathbf{v}_j^c$ are particles velocities in the centre of mass frame.\\
\indent Let us now re-write the last term in Eq. (\ref{angularconserv}) explicitly as function of the components of $\mathbf{v}_j^c$ and $\mathbf{r}_j$ and the rotation angle $\varphi_i$ as
\begin{eqnarray}
\sum_{j=1}^{N_i}\mathbf{r}_j\wedge\hat\mathbf{R}_{\varphi_i}\mathbf{v}^c_j=\cos\varphi_i\sum_{j=1}^{N_i}(x_jv_{yj}^c-y_jv_{xj}^c)+\nonumber\\
+\sin\varphi_i\sum_{j=1}^{N_i}(-x_jv_{xj}^c-y_jv_{yj}^c)=\nonumber\\
=\cos\varphi_i\sum_{j=1}^{N_i}\mathbf{r}_j\wedge\mathbf{v}_j^c-\sin\varphi_i\sum_{j=1}^{N_i}\mathbf{r}_j\cdot\mathbf{v}_j^c.
\end{eqnarray}
Therefore, one has
\begin{eqnarray}
\sum_{j=1}^{N_i}\mathbf{r}_j\wedge\mathbf{v}_j=N_i\mathbf{r}_i^c\wedge\mathbf{u}_i+\cos\varphi_i
\sum_{j=1}^{N_i}\mathbf{r}_j\wedge\mathbf{v}_j^c+\nonumber\\
-\sin\varphi_i\sum_{j=1}^{N_i}\mathbf{r}_j\wedge\mathbf{v}_j^c
\end{eqnarray}
and
\begin{eqnarray}
\sum_{j=1}^{N_i}\mathbf{r}_j\wedge(\mathbf{v}_j^c+\mathbf{u}_i)=\sum_{j=1}^{N_i}\mathbf{r}_j\wedge\mathbf{v}_j^c
+N_i\mathbf{r}_j\wedge\mathbf{u}_i.
\end{eqnarray}
Equating the two expressions above, and collecting the terms in sine and cosine finally leads to
\begin{equation}
(1-\cos\varphi_i)\sum_{j=1}^{N_i}\mathbf{r}_j\wedge\mathbf{v}_j^c+\sin\varphi_i\sum_{j=1}^{N_i}\mathbf{r}_j\cdot\mathbf{v}_j^c=0,
\end{equation}
that is verified when $\varphi_i$ is such that the definitions in Eqs (\ref{adef}-\ref{bdef}) hold, thus proving the conservation of the cell angular momentum for this choice of the rotation angle $\varphi_i$.\\
\indent Note that, the rotation operator allows only to preserve up to three conservation laws, therefore, in order to design an MPC scheme accounting for additional conservation laws (e.g., spin),  other suitable operators  should be introduced. Note also that, imposing the conservation of angular momentum with a rotation is possible only in two dimensions. However, in a 3D system it is still possible to conserve one of the three components of $\mathbf{L}=(L_x,L_y,L_z)$, say $L_z$, by imposing $z$ as rotation axis in each cell, and computing $\varphi_i$ from Eq.~(\ref{sincosphi}), where now $a_i$ is the $z$-component of the vector $\mathbf{a}_i=\sum_{j=1}^{N_i}\mathbf{r}_j \times (\mathbf{v}_j-\mathbf{u}_i)$.  
\bibliography{biblio.bib}
\end{document}